# Financial Trading Decisions based on Deep Fuzzy Self-Organizing Map

Pei Dehao[a], Luo Chao[a,b*]

( [a] School of Information Science and Engineering, Shandong Normal University, Jinan 250014, China

[b] Shandong Provincial Key Laboratory for Novel Distributed Computer Software Technology, Jinan 250014, China)

**Abstract:** The volatility features of financial data would considerably change in different periods, that is one of the main factors affecting the applications of machine learning in quantitative trading. Therefore, to effectively distinguish fluctuation patterns of financial markets can provide meaningful information for the trading decision. In this article, a novel intelligent trading system based on deep fuzzy self-organizing map (DFSOM) companied with GRU networks is proposed, where DFSOM is utilized for the clustering of financial data to acquire multiple fluctuation patterns in an unsupervised way. Firstly, in order to capture the trend features and evade the effect of high noises in financial data, the images of extended candlestick charts instead of raw data are processed and the obtained features are applied for the following unsupervised learning, where candlestick charts are produced with both price and volume information. Secondly, by using the candlestick features, a two-layer deep fuzzy self-organizing map is constructed to carry out the clustering, where two-layer models carry out the clustering in multiple time scales to improve the processing of time-dependent information. Thirdly, GRU networks are used to implement the prediction task, based on which an intelligent trading model is constructed. The feasibility and effectiveness of the proposed method are verified by using various real financial data sets.

## 1. Introduction

In recent years, with the development of information technology, quantitative trading gradually replaces the manual trading mode in the financial markets. Especially, companied with the rapid increment of artificial intelligence, more trading models are designed based on various intelligent algorithms, for instance, Czekalski et al [1] used the classical Feed Forward Multilayer Perceptron (ANN) to predict the FOREX trend. In [2], Lin et al. use a genetic algorithm (GA) to improve traditional technical analysis by learning the appropriate parameters for each trading rule, and the improved trading rules are passed through a novel neural network, the Echo State Network (ESN), which together give trading recommendations. Jia et al [3] proposed a LSTM-based agent to learn the temporal pattern in data and automatically trades according to the current market condition and the historical data. Therefore, machine learning has a broad application prospect in the field of quantitative trading.

---

[*]Corresponding author. E-mail: cluo79@gmail.com(C.Luo), 1134866100@qq.com(D.Pei)

The modeling and forecasting of financial time series have always been the important topics in the analysis of financial markets. Traditionally, statistical or stochastic process knowledge is used to model financial time series data, such as ARIMA [4]and GARCH [5]; However, the traditional statistical based modeling methods usually require the constraints of preset conditions, which are difficult to meet in the real environment. Therefore, various prediction methods based on intelligent algorithms have been proposed. For example, Tay et al [6] proposed the application of SVM to financial forecasting and investigated its functional characteristics in financial forecasting. Niu et al [7] propose a random data-time effective function and combine it with RBF neural network for financial time series forecasting. There are also neural networks such as RNN[8], LSTM [9], GRU [10], etc. These deep reinforcement learning based on recurrent neural networks can also extract useful features in complex financial data, and at the same time, their memory function can be used to learn the laws of data evolution in multidimensional financial derivatives. But the financial market, as a dynamical complex system, is vulnerable to the interference and influence of internal and external factors, which leads to the instability of financial time series and the change of fluctuation characteristics over time. This kind of market characteristics significantly affect the prediction effect of machine learning.

In order to capture more features in high dimensions, by means of image processing techniques, some researches have been implemented to learn time series in the form of images. For example, Wang et al [11] used methods such as Gramian angular sum/difference field (GASF/GADF) and Markov transition field (MTF) to encode time series into different types of images, and then used tied convolutional neural networks (TLED CNN) to classify the time series. Similarly, this idea spreads to the application of financial field. Barra et al [12] used the GAF imaging method in a trading system to encode financial time series data as images, a set of CNNs with the same structure but initialized with different kernel functions organized for image classification, and finally adopted a voting-based policy. Du et al [13] applied the continuous wavelet transform to the log return of financial time series and convert it to a greyscale wavelet transform spectrum, then built one shallow and one deep convolutional neural network (CNN) model and trained them with spectrum image as input to capture hidden patterns.

In terms of technical analysis of financial trading, the pattern recognition of k-line combinations is an important topic. However, In the traditional research of time series processing the high noises and uncertainty existing in the financial data evade the validity of sequence analysis. Therefore, whether financial time series can be studied from the perspective of images is an interesting problem. Cervelló-Royo et al [14] propose a new flag model that gives some weight matrix templates linking the IF-THEN rules for the identification patterns, apply the templates to the price windows to be checked, and decide whether to trade based on how well the K-line and the template match. Sezer et al [15] used 15 different technical metrics to convert financial time series into two-dimensional images and used 2-D CNNs to process the generated images. Candlestick charts is the most classical financial time series image, which contains abundant market information, and many researchers also study based on candlestick charts. For example, Tsai et al [16] extracted texture features of candlestick charts using Discrete Wavelet Transform, classified the features based on Euclidean distance, and predicted future stock fluctuations based on the categories; Hung et al [17] extracted the representation of candlestick charts using CNN-Autoencode and used 1D-CNN to predict price movements. But only price movement is present in candlestick charts. Volume is also a very important reference indicator for investments and, to our best knowledge, no one has integrated it into candlestick charts yet.

Self-organizing Map (SOM) is a kind of neural networks proposed by Kohonen[18]. However, different from the typical neural networks using error learning, SOM adopts competitive learning without the guidance of labels in an unsupervised way. During the processes of learning, the high-dimensional data can be mapped into the low-dimensional space on the premise of maintaining the original topological structures. Generally, SOM can be used as the unsupervised algorithm for clustering. For example, Kuo et al [19] applied SOM to the discovery and identification of stock patterns. Due to the excellent performance shown by deep learning in all directions, the structure of SOM has been motivated to become deeper. Rauber et al [20] proposed Growing Hierarchical Self-Organizing Map (GHSOM), which addresses the static structure of SOM models and the hierarchical relationship of data. Gharaee et al [21]proposed a hierarchical SOM-based online human behavior recognition system. Not only that, but SOM has also kept learning the ideas from other neural networks, like the most classical CNN. Liu et al [22] combined the ideas of convolutional neural network with SOM and proposed the deep self-organizing map (D-SOM).

Subsequently, Wicramasinghe et al [23] optimized the D-SOM model by proposing an extended version of the original DSOM (E-DSOM) and allowing a wider range of applications and faster training of the model.

Fuzzy logic[24][25] is integrated with many artificial neural networks [26] and is also introduced in some trading strategies [27]. In order to solve the problem of boundary vector swings during clustering, fuzzy logic was introduced to SOM. Mitra et al [28] proposed fuzzy self-organizing map (FSOM) in 1994 to handle fuzzy inputs and provide fuzzy classification. Hu et al [29] used FSOM for anomaly detection and activity prediction. Deng et al [30]applied FSOM to transmembrane segment detection. Miranda et al [31] used SOM to extract fuzzy rules. In the form of fuzzy rules, the relationship between variables in the system becomes clearer.

Motived by the existing works, in this article, a novel trading system based on deep fuzzy SOM is proposed. Firstly, the financial time series is transformed into the extended candlestick chart, where contains not only the price trend of the stock, but also the volume corresponding to each price of the stock, which provides us with richer information and features for image analysis. Secondly, the images are fed into the DFSOM model for cluster. Compared with the traditional DSOM, fuzzy technique is involved to handle with the uncertainty characteristics of financial time series data with high noises. Moreover, the Histogram of Oriented Gradient (HOG) descriptors instead of the pixels are implemented for the image processes, which can effectively extract the features of extended candlestick chart. Finally, GRU models are applied to carry out the prediction based on the results of clusters and the trading system can be achieved. The main contributions of this article are as follows.

1) Instead of raw data, the features of financial time series are learnt by using the images of extended candlestick charts, where the characteristics of $K$-line combination in each time window with additional volumes information can be fully captured.

2) In order to handle with high noises and large fluctuations of financial data, fuzzy technique is first introduced into DSOM, based on which the fluctuation patterns of financial behaviors can be effectively clustered.

3) To reveal the temporal-dependence features in different time scales, two-layer structure of DFSOM is constructed, where fluctuation patterns in multiple overlapping sliding windows can be learnt and fused.

The structure of the paper is as follows: the second part describes the proposed model in detail; the third part is model evaluation, in which various experiments are implemented to verify the usability of the proposed model; Finally, a conclusion is provided in the fourth part.

## 2. Methodology

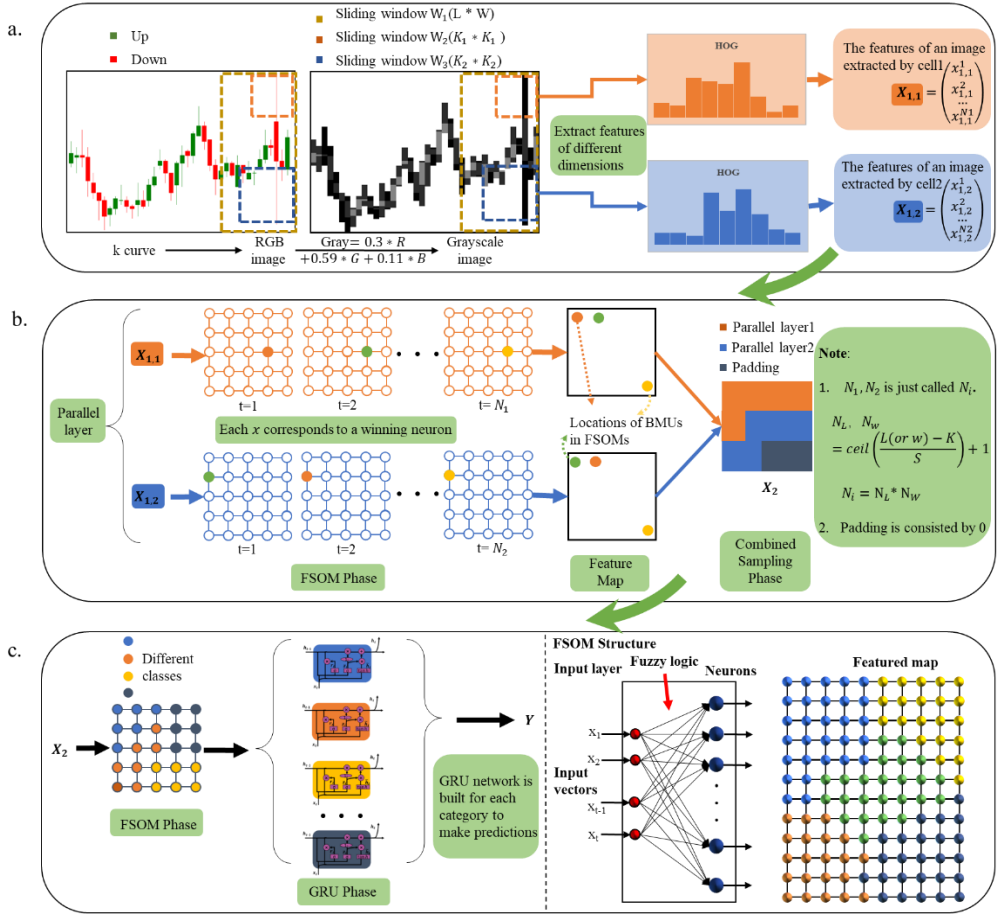

Fig.1.The model structure

Generally, the proposed model concludes three modules as shown in Fig.1. The first one is the feature extraction module, as shown in Fig.1.a. Sliding window $W_1$ is used to slide over the whole financial time series, and the data inside the window generates the extended candlestick chart. Among this sliding window $W_1$, there are also sliding windows $W_2$ and sliding window $W_3$ of different sizes, which are designed to extract HOG descriptors of different dimensions of the extended candlestick chart. The second is the classification module, as shown in Fig.1.b. The HOG descriptors extracted by sliding window $W_2$ and sliding window $W_3$ are used as the input of the parallel layer, which is composed of two FSOMs, and the FSOMs are used as the sampling layer. The number of features $N_1$ and $N_2$ of the image can be determined according to the size of the three sliding windows and the sliding step S. The HOG descriptors corresponding to the locations of the best match units (BMUs) in the FSOM are collected sequentially and stitched together, and the arithmetic square root of the number of spliced features needs to be an integer in order to facilitate the calculation. The third is the prediction module, as in Fig.1.c. This integrated feature

$X_2$ is used as the input of the third module, and after the last FSOM clustering, it is entered into different GRUs for prediction according to the results, and this FSOM layer has a larger receptive field than the FSOM layer in the second module. It is worth noting that there can be more than two parallel layers, here is just an example with two layers.

**2.1 Visibility graphs of financial time series**

In this section, the applied visibility graphs obtained from the financial time series are introduced. The traditional candlestick charts are constructed by using trading prices, i.e. the opening price, the highest price, the lowest price and closing price. However, the trading volumes are also the important information for decision making. Therefore, volumes are fused into candlestick charts for learning.

As we know, RGB image is made up of three color channels, i.e. red, green and blue. The traditional candlestick chart can be represented with red and green two colors, i.e. the red represents the fall and the green represents the rise. Here, we took advantage of the properties of the blue corresponding matrix. One minute of the data was used to synthesize the N minutes candlestick chart and assign the value of the volume of the price interval to the blue corresponding matrix, each of image contains a continuous ten candlestick charts. Because each minute corresponds to a different volume, the blue depth of each of the different locations is different, Fig.2.

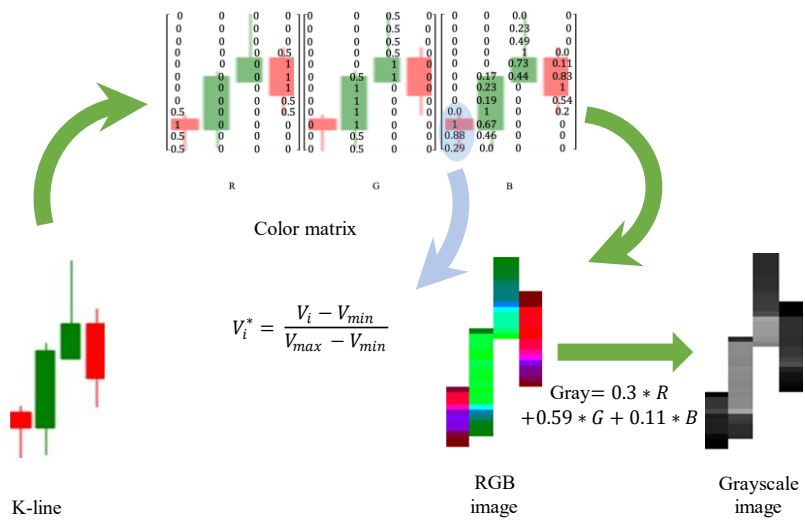

Fig.2 Changes in the process of the k-line in the processing process. So you can see in the RGB image that the blue depth of different prices in the candlestick chart is different.

Taking the synthesis of 30-minute candlestick chart as an example, and the algorithm is shown in Table 1. In the original data of financial time series, every minute data contains six items: trading time, opening price, highest price, lowest price, closing price, and volume. The first step is to normalize all the data and divide it into a list every 30 minutes and save these lists in a list X. Then initialize the three zero matrices R, G and B of size 100*10. In the second step, in chronological order, take one of the ten x from X. In the third step, for each x operation, each x contains 30 bars of minute data. The opening price, highest price, lowest price, closing price of this 30-minute data are calculated first, determine the color of the K-line, and then calculate the length of the upper shadow, lower shadow and body. In the case of a fall, we plot the K-line chart on the red matrix in the RGB image, and in the case of a rise, we plot the K-line chart on the green matrix. To distinguish between bodies and upper and lower shadows, we define the value corresponding to bodies to be 1 and the value corresponding to upper and lower shadows to be 0.5. In the fourth step, according to the highest and lowest price of each one-minute data, the position interval of this minute trading volume in the blue matrix is determined. The corresponding volume of this minute will be equally divided on this interval. When the 30 minutes of data traversal is completed, a K-line will be drawn. Since the blue matrix volume is superimposed, we have to normalize the volume of 30 minute. Return to step 3, until the ten candlestick charts are drawn. Return to step 2 to start the next image.

Table 1 Visibility graphs generation algorithm

| Algorithm I: GenerateImages() |
|---|
| Inputs: Minutes of the data |
| Output: All images with ten K-lines |
| 1: Normalize data |
| 2: for number of Inputs do |
| 3:     X ← Slice the Inputs every 30 minutes to a single list |
| 4: end for |
| 5: R、G、B ← Initialize to the zero matrix |
| 6: for number of X do |
| 7:     $x$ ← Get ten data points in order from X |
| 8:   for each epoch $x$ do |
| 9:     $index_x$ ← the location of x |
| 10:     open、high、low、close ← pick up four prices from each patch $x$ |
| 11:     if (open > close) |
| 12:       R [close : open, $index_x$] = 1 |
| 13:       R [open : high, $index_x$] = 0.5 |
| 14:       R [low : close, $index_x$] = 0.5 |
| 15:     else |
| 16:       G [open : close, $index_x$] = 1 |
| 17:       G [close : high, $index_x$] = 0.5 |
| 18:       G [low : open, $index_x$] = 0.5 |
| 19:     end if |
| 20:     for each epoch $x[index_x]$ do |
| 21:       $k$ ← Get per minute data from $x[index_x]$ |
| 22:       B [$k_{low} : k_{high}, index_x$] = volume/($k_{high} - k_{low}$) |
| 23:       normalize B [$k_{low} : k_{high}, index_x$] |
| 24:     end for |
| 25:   end for |
| 26: end for |

## 2.2 The Histogram of Orientation Gradients (HOG) based feature extraction

Different from the existing works, the obtained visibility graphs are not directly processed using pixel values. In order to reduce redundant information, histogram of orientation gradients is applied, which can extract distinguishing information from an image and discards irrelevant information to simplify the image representation. The process of extracting HOG from an image is shown in Fig.1. a. Each Window is given a HOG description and the proposed model takes each HOG description as DFSOM input.

For the calculation of HOG, RGB components should be converted to grayscale images as follows:

$$\text{Gray} = 0.3*R + 0.59*G + 0.11*B \tag{1}$$

For the image, one can construct two filters, $[-1 \quad 0 \quad 1]$ for the horizontal direction and $[-1 \quad 0 \quad 1]^T$ for the vertical direction, which can be used to convolve the image horizontally and vertically. Then, the gradient $G_x$ of the image in the horizontal direction and the gradient $G_y$ of the image in the vertical direction can be calculated as follows:

$$G_x(x, y) = I(x+1, y) - I(x-1, y) \tag{2}$$

$$G_y(x, y) = I(x, y+1) - I(x, y-1) \tag{3}$$

Subsequently, one need to calculate the orientation of the point, starting with the tangent angle formed by the gradient.

$$\theta(x, y) = \arctan\left(\frac{G_y(x, y)}{G_x(x, y)}\right) \tag{4}$$

Let's take $\theta$ from 0° to 180° and divide $\theta$ into $K$ intervals. It is assumed that $K$ equals to 9, then the span of each interval $\Delta$ is 20°, and the orientation records which interval the tangent angle enters can be calculated:

$$Ori = \left\lfloor \frac{\theta}{\Delta} \right\rfloor \tag{5}$$

Since the value of the orientation belongs to $(0,1,\ldots,8)$, HOG can be calculated by counting the orientation of each pixel.

As shown in Fig.1.a, in order to capture the fluctuant patterns in different time scales, two overlapping sliding windows $W_2$ and $W_3$ with different sizes are used, where the HOG features of two time segments are calculated separately as the inputs of DFSOM. Here, the number of obtained features should be calculated as follows:

$$N_L, N_W = \text{ceil}\left(\frac{L(\text{or } W) - K}{S}\right) + 1 \tag{6}$$

$$N_i = N_L \times N_W \tag{7}$$

where ceil is a function to find the upper bound minimum integer, $N_L$ and $N_W$ represent the number of features in the horizontal and vertical directions, L and W are the length and width of sliding windows $W_1$, K represents the side length of $W_2$ and $W_3$, S represents the sliding step of $W_2$ and $W_3$, and $N_i$ is the total number of features.

**2.3 The construction of Deep Fuzzy Self-Organization Map**

In this section, two-layer DFSOM is constructed to learn the obtained features $X_{1,1}$ and $X_{1,2}$ from time segments with distinct time scales. The structure of DFSOM is shown in Fig.3, where the inputs of the network are the feature vectors extracted by each sliding unit in the first module, just like the convolution layer of CNN. DFSOM uses the SOM neural network as a pooling layer, which serves the purpose of dimensionality reduction and extraction of higher dimensional features, while collecting the winning neurons generated by SOM in a sequential manner. The parallel layer serves to obtain multi-dimensional features of the images, with the aim of improving the accuracy of the clustering. Finally, the features of the whole picture collected by the parallel layer are linearly stitched together. For the convenience of subsequent calculation, if the spliced features cannot regenerate a square, 0 needs to be added after these features to regenerate a feature matrix with a square shape. This square feature $X_2$ is the input of the third module, and the resulting output is the clustering of the image, as in Fig.1.c.

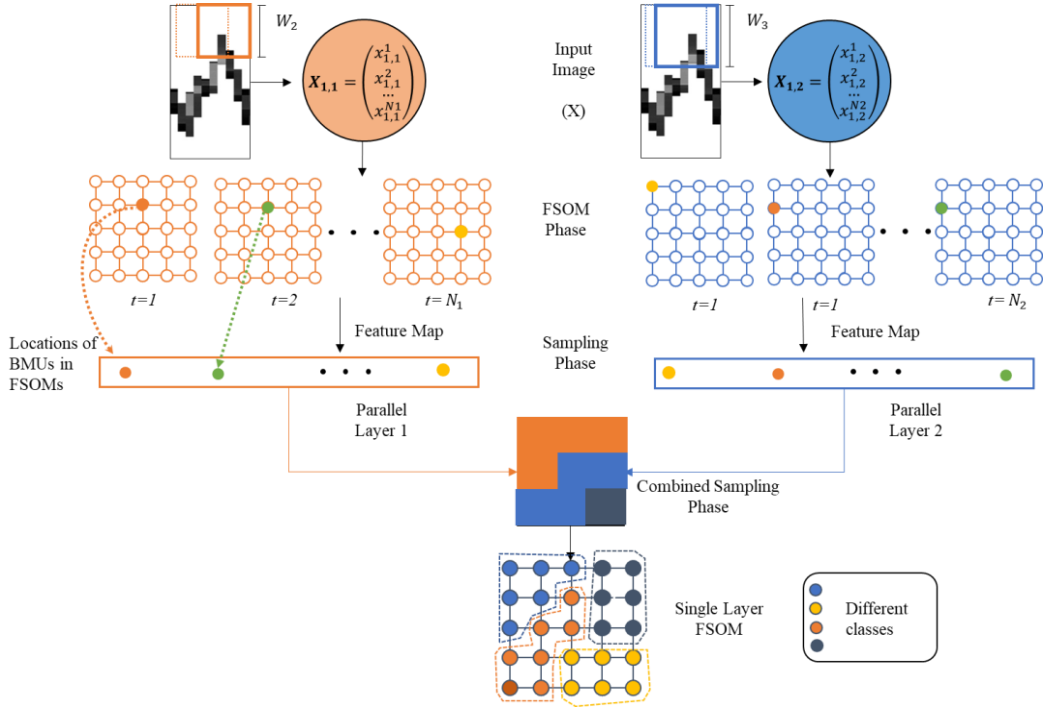

Fig.3 DFSOM structure

Table 2 DFSOM algorithm

| Algorithm II: DFSOM Training |
|---|
| Inputs: Training set of images(X) |
| Output: Trained DFSOM |
| 1:     Random Weight initialization |
| 2:     **for** each epoch e do |
| 3:         **for** number of training samples do |
| 4:             $x \leftarrow$ pick random input record from X |
| 5:             **for** each Layer II do |
| 6:                 *featureMapList* $\leftarrow$ empty list of length P |
| 7:                 **for** each parallel FSOM layer p do |
| 8:                       *featureMapList*[p]$\leftarrow$*parallelLayer*($x$) |
| 9:                 **end for** |
| 10:                $x \leftarrow$ *CombinedSampling(featureMapList)* |
| 11:             **end for** |
| 12:             OutputSOM $\leftarrow$ Algorithm III ($x$) |
|                 % BMU is the position where the weight is the least distance from x |
| 13:        **end for** |
| 14: **end for** |
| Procedure I: ParallelLayer |
| Inputs: Input record($x$), Number of patches(p) |
| Outputs: Sampled *featureMap* |
| 1:     *featureMap* $\leftarrow$ empty list of length p |
| 2:     **for** each patch $\acute{x}$ do: |
| 3:         $index_x \leftarrow$ the location of $\acute{x}$ w.r.t $x$ |
| 4:         $BMU_{\acute{x}} \leftarrow$ get BMU index for $\acute{x}$ on corresponding $FSOM_{\acute{x}}$ |
| 5:         *featureMap[index]* $\leftarrow$ $BMU_{\acute{x}}$ |
| 6:     **end for** |
| Procedure II: CombinedSampling |
| Inputs: List of feature maps from each parallel layer(*featureMapList*) |
| Outputs: Combined Feature Map |
| 1:     *ComFeatureMap* $\leftarrow$ Append *featureMapList* to a single list |
| 2:     $l \leftarrow$ length of *comFeatureMap* |
| 3:     **if** $\sqrt{l} \notin N$ where N={1,2,3,4,…} |
| 4:         Use zero-padding on comFeatureMap until $\sqrt{l} \in N$ |
| 5:     CFM $\leftarrow$ Reshape *comFeatureMap* to a 2D vector of size $\sqrt{l} * \sqrt{l}$ |
| 6:     **return** CFM |

FSOM algorithm is used in the FSOM layer of the whole model, the FSOM neural network uses a two-dimensional network structure as shown in Fig.1.c and the algorithm is shown in Table 3. Let M denote the number of input samples, N the number of input vector components, and K the number of output neurons. The learning algorithm consists of the following steps.

Step 1)    Randomize the initial values of the components of the weight vectors.

Step 2)    Input all samples $X_l = [X_{l,1}, X_{l,2}, \ldots, X_{l,N}], l = 1,2,\ldots,M$.

Step 3)    Calculate the Euclidean distances from each sample $X_l$ to all output neurons

$$d_{lj}(t) = \sqrt{\sum_{i=1}^{N}(X_{li} - W_{ij}(t))^2}, l=1,2,\ldots,M, j=1,2,\ldots,K. \tag{8}$$

Step 4)    Compute the memberships of each sample to all neurons

$$R_{ij}(t) = \frac{\frac{1}{d_{lj}^2(t)}}{\sum_{m=1}^{K}\left(\frac{1}{d_{lm}^2(t)}\right)}, l=1,2,\ldots,M, j=1,2,\ldots,K. \qquad (9)$$

Step 5) Adjust the weights of each neuron according to the computed memberships

$$W_{ij}(t+1) = W_{ij}(t) + \frac{\sum_{l=1}^{M} R_{lj}(t) \cdot (X_{li} - W_{ij}(t))}{\sum_{l=1}^{M} R_{lj}(t)} \qquad (10)$$

Step 6) Determine the stability condition of the network

$$\max_{\substack{1<i<N \\ 1\leq j\leq K}} \{\|W_{ij}(t+1) - W_{ij}(t)\|\} < \varepsilon. \qquad (11)$$

After DFSOM clustering, images with similar features converge into one category, and we set up a GRU network for each category. Through the prediction of GRU network, we completed the prediction of time series.

**Table 3** FSOM algorithm

| Algorithm III: FSOM Training |
|---|
| Inputs: Training set of images(X) |
| Outputs: Trained SOM |
| 1:     Random Weight initialization |
| 2:     $\varepsilon \leftarrow$ Termination conditions |
| 3:     **While** $\max(|W_{i,j}(t+1) - W_{i,j}(t)|) \leq \varepsilon$: |
| 4:         **for** number of neurons in SOM do |
| 5:             **for** number of training samples do |
| 6:                 $x \leftarrow$ pick random input record from X |
| 7:                 $d \leftarrow$ Calculate the Euclidean distance from each $x$ to $w$ |
| 8:             **end for** |
| 9:         $R \leftarrow$ Calculate the membership of each input vector to this neuron |
| 10:        $W_i \leftarrow W_i + \Delta W_i \leftarrow$ Adjust neurons according to membership |
| 11:    **end for** |

## 3 Experiments

**Datasets:** Two kinds of real financial datasets are utilized. One is China's commodity futures dataset in 2020, and the other is foreign exchange dataset in 2016. Due to the large volume of experiments, some windows are selected for detailed discussion in the subsequent section, including AG (Silver) and EG (Ethylene Glycol) in the commodity futures dataset, and AUDJPY and EURUSD in the foreign exchange. The Appendix section presents the experimental results for the entire dataset.

**Parameters:** To ensure the comparability of the experimental results, the same parameters are used for the same modules. For example, for the models of EDSOM and DFSOM mentioned in subsection 3.1, both using the architecture given in subsection 2.3. The sliding window sizes of the two extracted features are 3×3 and 6×6, and the corresponding sliding steps are 1 and 2. The network structure of parallel SOM sampling layers both are 15×15, and the last SOM network structure responsible for classification is 8×8. The weights of both SOM and FSOM use the same random initialization method.

**Fixed fee:** Since futures trading fees fluctuate and are different from each platform, for statistical convenience, we set the total transaction costs for each commodity futures to 0.2% including 0.1% for explicit transaction costs (commission fee and stamp duty, etc.) and 0.1% for implicit transaction costs (slippage). Similarly, allowing for the slippage in Forex, we set the total transaction cost for each type of Forex to 0.1%.

### 3.1 Model comparison and evaluation

In order to verify the performance of the model, five additional trading models are used to carry out comparative studies. This includes comparisons with Model1 and Model 2 which were predicted using the raw data approach, as well as ablation experiments performed with Model 3, Model 4,and Model 6, and comparative experiments with model5.

Model 1)   A single BP is used in the prediction module;

Model 2)   A single GRU is used as the prediction module;

Model 3)   HOG is used to extract features, EDSOM is used for clustering, and GRU is used for prediction;

Model 4)   DFSOM is used for clustering and GRU for prediction;

Model 5)   HOG is used to extract features, DFSOM is used for clustering, and BP is used for prediction;

Model 6)   The proposed model. HOG is used to extract features, DFSOM is used for clustering, and GRU is used for prediction.

The following performance indicators will be used in experiments to evaluate the performance of the models:

1) PR: Profit rate

2) TN: The total number of trades

3) TN+: The number of profitable trades

4) TN-: The number of non-profitable trades

5) Avg_return: The average return of all trades:

$$Avg\_return = \frac{PR}{TN} * 100$$

6) Avg_profit: The average profits of the profitable trades:

$$Avg\_profit = \frac{Total\_profits}{TN+} * 100$$

7) Avg_loss: The average loss of the non-profitable trades:

$$Avg\_loss = \frac{Total\_loss}{TN-} * 100$$

8) P/L: The profit and loss ratio.

9) Accuracy: The decision accuracy.

$$Accuracy = \frac{TN+}{TN}$$

## 3.2 Trading strategies

Based on the proposed prediction model, an intelligent trading system can be constructed. Generally, the trading strategy is concise, that is, going long when forecasting a rise in the coming time window or selling short for a predicting falling. However, there are some trivial fluctuations should be removed. Therefore, as shown in Fig.4, two thresholds, i.e. long threshold LT and short threshold ST, are set for the trading strategy. The thresholds are determined based on the closing price at time $t$ multiplied with a fix rate. When the financial price is predicted for a significant rising (exceeding the threshold LT) at the end of the next time window, a buying-long operation is performed using the price at time $t+1$ and closing the position at the end of the next time window. Likewise, the selling-short operation can be carried out.

To sum up, the forecast price at time $t+1$ is assumed to be $P_{t+1}$, the trading strategies is as follows:

a. $P_{t+1} > LT$, buy long;

b. $P_{t+1} < ST$, sell short;

c. $ST < P_{t+1} < LT$, no operation.

d. Position will be held in a time period.

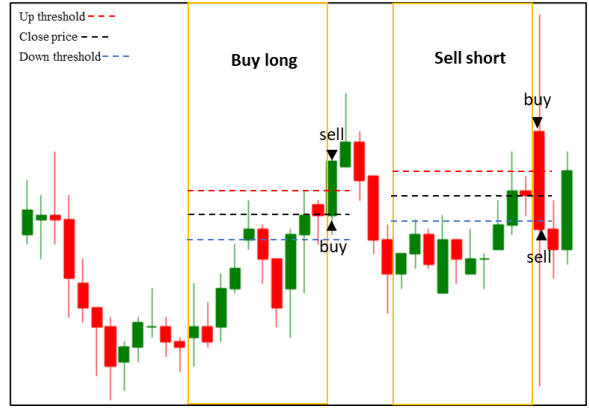

Fig.4 Trading Rules

**3.3 Experiments on Commodity Futures**

Here, the 30-minute data of the commodity futures is used, i.e., time window is set to be 30 minutes. According to the trading strategy, when the closing prices of commodity futures after 30 minutes are predicted to rise above the threshold LT, the long operation will be implemented, and the position will be held in the coming 30 minutes. Similarly, selling-short operation will be implemented when the closing prices are forecasted to fall below the threshold ST. However, when the forecasting prices belongs to the range between the thresholds LT and ST, there is no trading action. In the following, the commodity futures AG and EB are used for discussion. The time windows of the training and test sets are listed in Table 4.

Table 4 The time window of futures experimental data

| Sliding windows | Training_period | Testing_period |
| --- | --- | --- |
| 1 | 2020.05.18-2020.08.28 | 2020.08.31-2020.09.11 |
| 2 | 2020.05.18-2020.09.11 | 2020.09.14-2020.09.25 |
| 3 | 2020.06.01-2020.09.25 | 2020.09.28-2020.10.12 |
| 4 | 2020.06.15-2020.10.12 | 2020.10.13-2020.10.27 |
| 5 | 2020.06.30-2020.10.27 | 2020.10.28-2020.11.11 |
| 6 | 2020.07.15-2020.11.11 | 2020.11.12-2020.11.26 |

Profitability and stability are the two most important indicators to evaluate a trading system. If a system is not stable enough, fluctuates a lot, or is not profitable, then the trading system cannot be used in practice. Figures 5 and 6 show the details of the profitability of futures AG and EB. The bar chart shows the profit ratio for each sliding window and the line chart shows the accumulation of profits over the sliding window. From the figure, we can see that the profit ratio of Model 6 is the least volatile and the most stable compared to the other five, and the accumulated profit is the highest. During the three months of the test set, AG's accumulated profit was 26.34% and EB's accumulated profit was 10.03%.

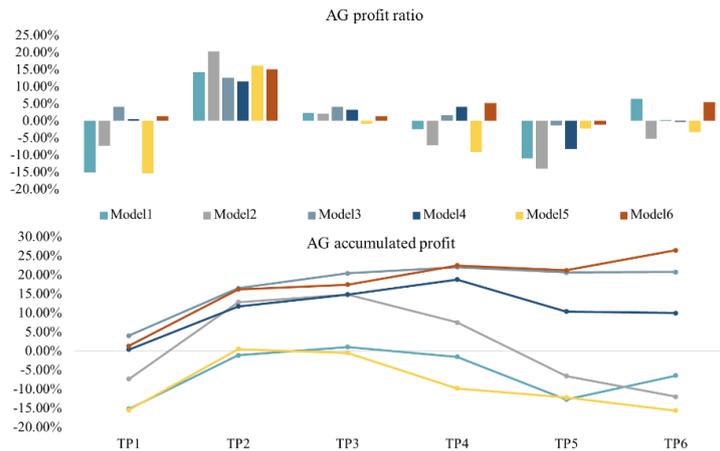

Fig 5 In futures AG, the profit rate and cumulative return of 6 models in 6 periods respectively

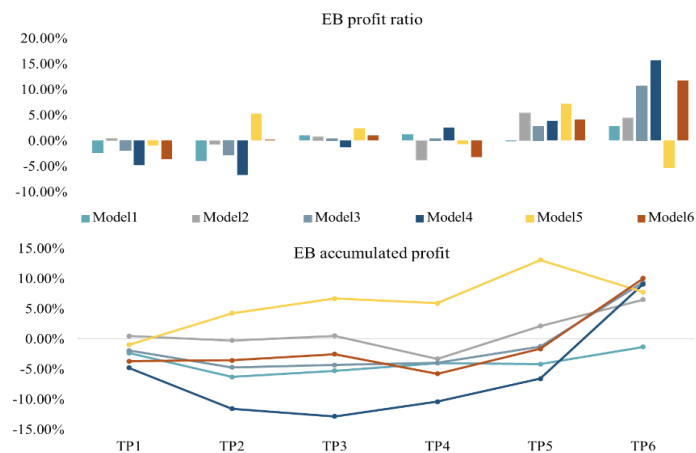

Fig 6 In futures EB, the profit rate and cumulative return of 6 models in 6 periods respectively

The P/L is an important indicator of the profitability of a trading system. A successful trade should have a P/L value bigger than 1, and the bigger the ratio, the greater the possibility of profitability. As can be seen in Figures 7 and 8, Model 6 has an average P/L ratio greater than 1 in 2/3 of the cases, and in EB, the P/L ratio is as high as 1.764.

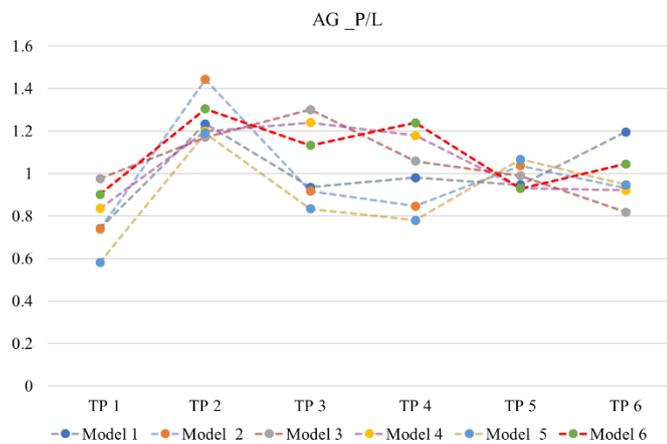

Fig 7 P/L corresponding to future AG in 6 test periods

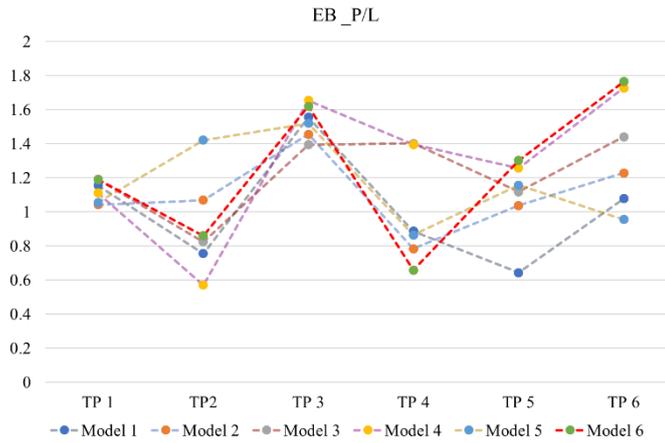

Fig 8 P/L corresponding to future eb in 6 test periods

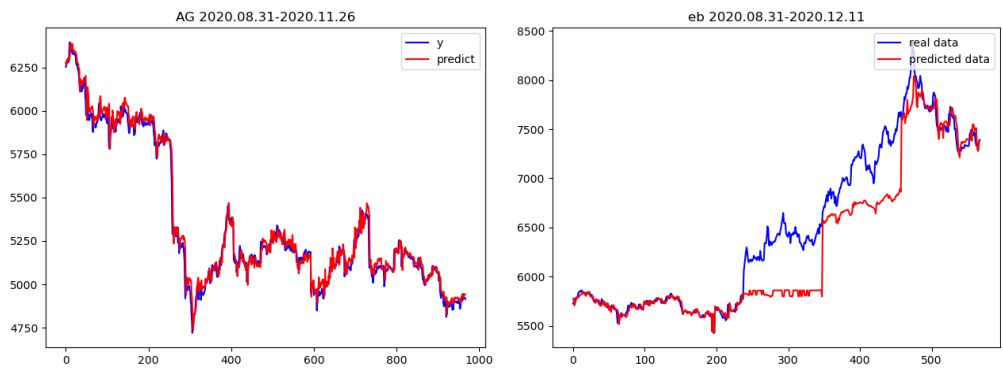

Fig 9 Total fitting curves of future AG and future EB in 6 test periods

The fitting curves are shown in Fig.9. Tables 5 and 6 show the details of the trades where the six different trading models are evaluated by the 9-evaluation metrics shown in Section 3.2. As can be seen from tables, Model 6 can get more profits with fewer transactions compared to models 1 and 2. Not only does it require less training time than Model 3 and 4, but it also performs better in terms of yield and accuracy. Compared to Model 5, we can see from the comparison of all aspects that the GRU prediction model is better than the BP prediction model with DFSOM clustering model is a better fit.

Table 5 nine performance metrics comparisons of future AG with 6 trading models, where TP is the corresponding Testing_period in Table 4

| M | TP | PR | SUMPR | TN | TN+ | TN- | Avg_return | Avg_profit | Avg_loss | P/L | Accuracy |
|---|----|-----|-------|-----|-----|-----|------------|------------|----------|-------|----------|
| 1 | 1 | -15.29% | -6.52% | 168 | 74 | 94 | -0.091 | 0.290 | -0.391 | 0.742 | 44.05% |
|   | 2 | 14.06% |       | 168 | 92 | 76 | 0.084 | 0.464 | -0.376 | 1.234 | 54.76% |
|   | 3 | 2.19% |        | 67  | 38 | 29 | 0.033 | 0.314 | -0.336 | 0.935 | 56.72% |
|   | 4 | -2.57% |       | 186 | 89 | 97 | -0.014 | 0.264 | -0.269 | 0.981 | 47.85% |
|   | 5 | -11.22% |      | 186 | 80 | 106 | -0.060 | 0.351 | -0.371 | 0.946 | 43.01% |
|   | 6 | 6.31% |        | 186 | 98 | 88 | 0.034 | 0.257 | -0.215 | 1.195 | 52.69% |
| 2 | 1 | -7.46% | -12.07% | 168 | 86 | 82 | -0.044 | 0.295 | -0.400 | 0.738 | 51.19% |
|   | 2 | 20.19% |       | 168 | 93 | 75 | **0.120** | 0.492 | -0.341 | **1.443** | 55.36% |
|   | 3 | 1.99% |        | 67  | 38 | 29 | 0.030 | 0.311 | -0.339 | 0.917 | 56.72% |

| | | | | | | | | | | |
|---|---|---|---|---|---|---|---|---|---|---|
| | 4 | -7.29% | | 186 | 87 | 99 | -0.039 | 0.243 | -0.287 | 0.847 | 46.77% |
| | 5 | -14.1% | | 186 | 72 | 114 | -0.076 | 0.370 | -0.357 | 1.036 | 38.71% |
| | 6 | -5.4% | | 186 | 85 | 101 | -0.029 | 0.228 | -0.245 | 0.931 | 45.70% |
| 3 | 1 | 3.92% | 20.63% | 132 | 72 | 60 | 0.030 | 0.372 | -0.381 | 0.976 | 54.55% |
| | 2 | 12.44% | | 142 | 80 | 62 | 0.088 | 0.460 | -0.393 | 1.170 | 56.34% |
| | 3 | 3.96% | | 61 | 33 | 28 | 0.065 | 0.346 | -0.266 | 1.301 | 54.10% |
| | 4 | 1.62% | | 142 | 72 | 70 | 0.011 | 0.272 | -0.257 | 1.058 | 50.70% |
| | 5 | -1.42% | | 159 | 78 | 81 | -0.009 | 0.374 | -0.378 | 0.989 | 49.06% |
| | 6 | 0.11% | | 147 | 81 | 66 | 0.001 | 0.239 | -0.292 | 0.818 | 55.10% |
| 4 | 1 | 0.27% | | 159 | 87 | 72 | 0.002 | 0.324 | -0.388 | 0.836 | 54.72% |
| | 2 | 11.37% | 9.90% | 158 | 85 | 73 | 0.072 | 0.472 | -0.393 | 1.199 | 53.80% |
| | 3 | 3.10% | | 63 | 33 | 30 | 0.049 | 0.352 | -0.284 | 1.240 | 52.38% |
| | 4 | 3.93% | | 174 | 87 | 87 | 0.023 | 0.297 | -0.252 | 1.179 | 50.00% |
| | 5 | -8.37% | | 168 | 76 | 92 | -0.050 | 0.365 | -0.392 | 0.930 | 45.24% |
| | 6 | -0.40% | | 161 | 83 | 78 | -0.002 | 0.238 | -0.259 | 0.921 | 51.55% |
| 5 | 1 | -15.58% | -15.75% | 129 | 59 | 70 | -0.121 | 0.255 | **-0.438** | 0.582 | 45.74% |
| | 2 | 16% | | 145 | 84 | 61 | 0.110 | 0.490 | -0.412 | 1.189 | **57.93%** |
| | 3 | -1% | | 56 | 29 | 27 | -0.018 | 0.301 | -0.361 | 0.834 | 51.79% |
| | 4 | -9.34% | | 151 | 67 | 84 | -0.062 | 0.230 | -0.295 | 0.780 | 44.37% |
| | 5 | -2.41% | | 139 | 64 | 75 | -0.017 | 0.383 | -0.359 | 1.067 | 46.04% |
| | 6 | -3.42% | | 123 | 56 | 67 | -0.028 | 0.233 | -0.246 | 0.947 | 45.53% |
| 6 | 1 | 1.23% | **26.34%** | 130 | 70 | 60 | 0.009 | 0.347 | -0.385 | 0.901 | 53.85% |
| | 2 | 14.88% | | 144 | 79 | 65 | 0.103 | **0.510** | -0.391 | 1.304 | 54.86% |
| | 3 | 1.22% | | 62 | 31 | 31 | 0.020 | 0.343 | -0.303 | 1.132 | 50.00% |
| | 4 | 5.04% | | 158 | 80 | 78 | 0.032 | 0.296 | -0.239 | 1.238 | 50.63% |
| | 5 | -1.29% | | 154 | 78 | 76 | -0.008 | 0.352 | -0.379 | 0.929 | 50.65% |
| | 6 | 5.26% | | 136 | 78 | 58 | 0.039 | 0.232 | -0.222 | 1.045 | 57.35% |

Table 6 nine performance metrics comparisons of future EB with 6 trading models

| M | TP | PR | SUMPR | TN | TN+ | TN- | Avg_return | Avg_profit | Avg_loss | P/L | Accuracy |
|---|---|---|---|---|---|---|---|---|---|---|---|
| 1 | 1 | -2.37% | -1.34% | 78 | 32 | 46 | -0.030 | 0.303 | -0.263 | 1.156 | 41.03% |
| | 2 | -3.95% | | 80 | 39 | 41 | -0.049 | 0.258 | -0.342 | 0.755 | 48.75% |
| | 3 | 1.03% | | 32 | 14 | 18 | 0.032 | 0.423 | -0.272 | 1.555 | 43.75% |
| | 4 | 1.28% | | 30 | 17 | 13 | 0.043 | 0.550 | -0.620 | 0.886 | 56.67% |
| | 5 | -0.17% | | 38 | 23 | 15 | -0.004 | 0.529 | -0.822 | 0.643 | **60.53%** |
| | 6 | 2.84% | | 69 | 36 | 33 | 0.041 | 0.527 | -0.489 | 1.078 | 52.17% |
| 2 | 1 | 0.45% | 6.51% | 84 | 42 | 42 | 0.005 | 0.266 | -0.255 | 1.042 | 50.00% |
| | 2 | -0.71% | | 69 | 32 | 37 | -0.010 | 0.273 | -0.255 | 1.069 | 46.38% |
| | 3 | 0.76% | | 34 | 15 | 19 | 0.022 | 0.390 | -0.268 | 1.456 | 44.12% |
| | 4 | -3.80% | | 13 | 5 | 8 | -0.292 | 0.730 | -0.931 | 0.784 | 38.46% |
| | 5 | 5.44% | | 78 | 43 | 35 | 0.070 | 0.587 | -0.566 | 1.037 | 55.13% |
| | 6 | 4.38% | | 78 | 39 | 39 | 0.056 | 0.604 | -0.492 | 1.228 | 50.00% |
| 3 | 1 | -1.92% | 9.38% | 71 | 29 | 42 | -0.027 | 0.306 | -0.257 | 1.191 | 40.85% |
| | 2 | -2.79% | | 63 | 30 | 33 | -0.044 | 0.277 | -0.336 | 0.824 | 47.62% |
| | 3 | 0.37% | | 25 | 11 | 14 | 0.015 | 0.386 | -0.277 | 1.394 | 44.00% |
| | 4 | 0.35% | | 28 | 12 | 16 | 0.013 | 0.588 | -0.420 | 1.402 | 42.86% |
| | 5 | 2.72% | | 67 | 34 | 33 | 0.041 | 0.607 | -0.543 | 1.118 | 50.75% |
| | 6 | 10.65% | | 88 | 46 | 42 | 0.121 | 0.632 | -0.439 | 1.44 | 52.27% |

| | | | | | | | | | | |
|---|---|---|---|---|---|---|---|---|---|---|
| 4 | 1 | -4.81% | 9.08% | 78 | 28 | 50 | -0.062 | 0.282 | -0.254 | 1.110 | 35.90% |
| | 2 | -6.76% | | 44 | 17 | 27 | -0.154 | 0.223 | -0.391 | 0.570 | 38.64% |
| | 3 | -1.29% | | 29 | 9 | 20 | -0.044 | 0.416 | -0.251 | 1.654 | 31.03% |
| | 4 | 2.46% | | 19 | 11 | 8 | 0.129 | 0.466 | -0.334 | 1.395 | 57.89% |
| | 5 | 3.82% | | 62 | 31 | 31 | 0.062 | 0.603 | -0.480 | 1.257 | 50.00% |
| | 6 | 15.65% | | 90 | 47 | 43 | **0.174** | **0.708** | -0.410 | 1.727 | 52.22% |
| 5 | 1 | -0.97% | 7.75% | 88 | 41 | 47 | -0.011 | 0.272 | -0.258 | 1.055 | 46.59% |
| | 2 | 5.21% | | 73 | 39 | 34 | 0.071 | 0.345 | -0.243 | 1.421 | 53.42% |
| | 3 | 2.44% | | 29 | 15 | 14 | 0.084 | 0.421 | -0.277 | 1.519 | 51.72% |
| | 4 | -0.76% | | 74 | 39 | 35 | -0.010 | 0.494 | -0.573 | 0.864 | 52.70% |
| | 5 | 7.16% | | 79 | 44 | 35 | 0.091 | 0.521 | -0.450 | 1.157 | 55.70% |
| | 6 | -5.33% | | 86 | 39 | 47 | -0.062 | 0.523 | -0.548 | 0.956 | 45.35% |
| 6 | 1 | -3.69% | **10.03%** | 78 | 29 | 49 | -0.047 | 0.303 | -0.255 | 1.190 | 37.18% |
| | 2 | 0.13% | | 74 | 40 | 34 | 0.002 | 0.283 | -0.329 | 0.860 | 54.05% |
| | 3 | 1.03% | | 32 | 14 | 18 | 0.032 | 0.359 | **-0.221** | 1.619 | 43.75% |
| | 4 | -3.24% | | 28 | 12 | 16 | -0.116 | 0.263 | -0.400 | 0.658 | 42.86% |
| | 5 | 4.10% | | 67 | 33 | 34 | 0.061 | 0.597 | -0.458 | 1.302 | 49.25% |
| | 6 | 11.7% | | 83 | 41 | 42 | 0.141 | 0.681 | -0.386 | **1.764** | 49.40% |

### 3.4 Experiments on Forex

Forex trading follows the same rules as futures trading, and we adopt basically the same trading strategy. The only difference is that we chose to use 60-minute data for commodity forex instead of 30-minute data, because there is too much minute data in one day's forex data. The trading strategy is the same as for futures trading, when the closing prices of commodity futures after 60 minutes are predicted to rise above the threshold LT, the long operation will be implemented, and the position will be held in the coming 60 minutes. Similarly, selling-short operation will be implemented when the closing prices are forecasted to fall below the threshold ST. However, when the forecasting prices belongs to the range between the thresholds LT and ST, there is no trading action. The profit ratio for trading is a cumulative percentage increase. In the following, the forex AUDJPY and EURCAD are used for discussion. The time windows of the training and test sets are listed in Table 7.

Table 7 The time window of futures experimental data

| Sliding windows | Training_period | Testing_period |
|---|---|---|
| 1 | 2016.01.03-2016.04.01 | 2016.04.03-2016.04.17 |
| 2 | 2016.01.18-2016.04.17 | 2016.04.18-2016.05.02 |
| 3 | 2016.02.02-2016.05.02 | 2016.05.03-2016.05.17 |
| 4 | 2016.02.17-2016.05.17 | 2016.05.18-2016.06.01 |
| 5 | 2016.03.03-2016.06.01 | 2016.06.02-2016.06.16 |
| 6 | 2016.03.20-2016.06.16 | 2016.06.17-2016.07.01 |

Figures 10 and 11 show the details of the profitability of forex AUDJPY and EURCAD. The bar chart shows the profit ratio for each sliding window and the line chart shows the accumulation of profits over the sliding window. Those figures show that the profit ratio of Model 6 is always positive, especially for AUDJPY at TP4 and EURCAD at TP2 and TP3, while the other five models are negative, our proposed Model 6 is still profitable. In the period 2016.04.03 to 2020.07.01, Model 6 had the highest accumulated profit on both Forex, with an accumulated profit of 8.81% on AUDJPY and 5.65% on EURCAD.

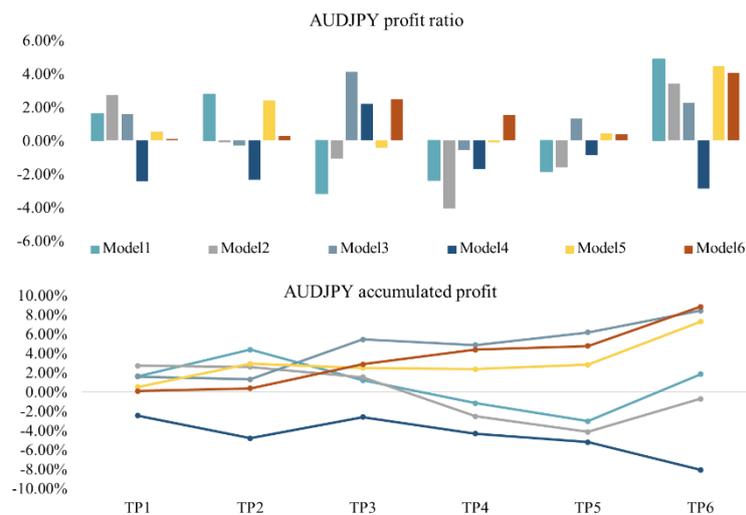

Fig 10 In forex AUDJPY, the profit rate and cumulative return of 6 models in 6 periods respectively

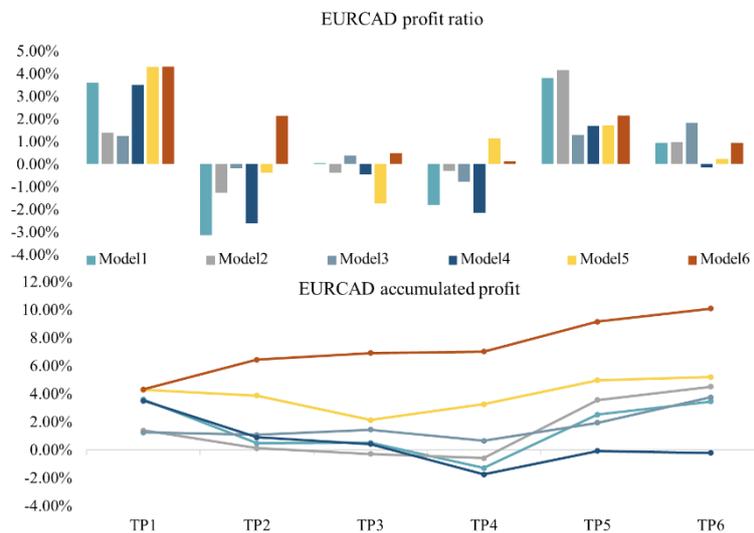

Fig 11 In forex EURCAD, the profit rate and cumulative return of 6 models in 6 periods respectively

As can be seen in Figures 12 and 13, Model 6 has an average P/L ratio greater than 1 in 5/6 of the cases, and in AUDJPY, the P/L ratio is as high as 1.441.

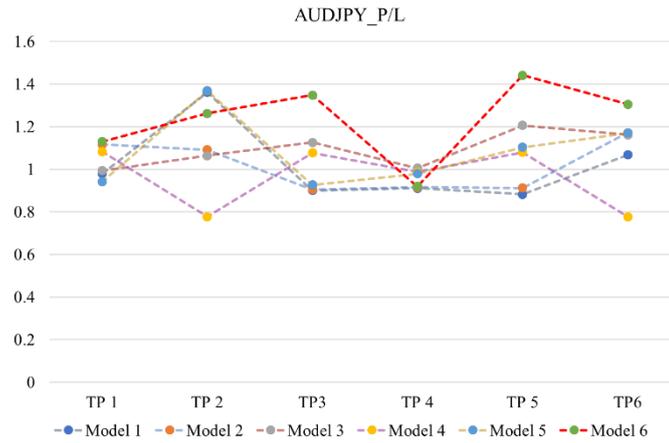

Fig 12 P/L corresponding to Forex AUDJPY in 6 test periods

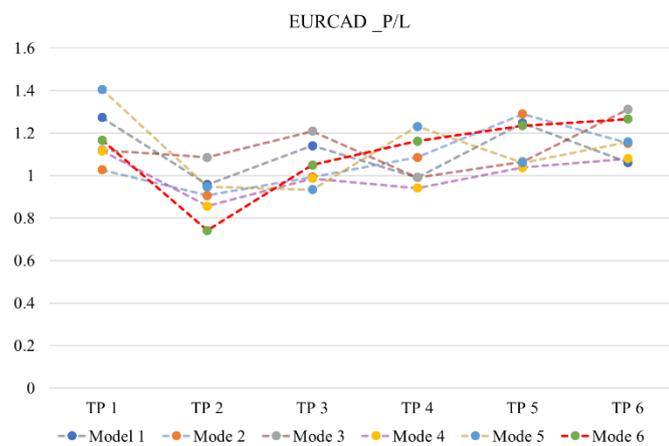

Fig 13 P/L corresponding to Forex EURCAD in 6 test periods

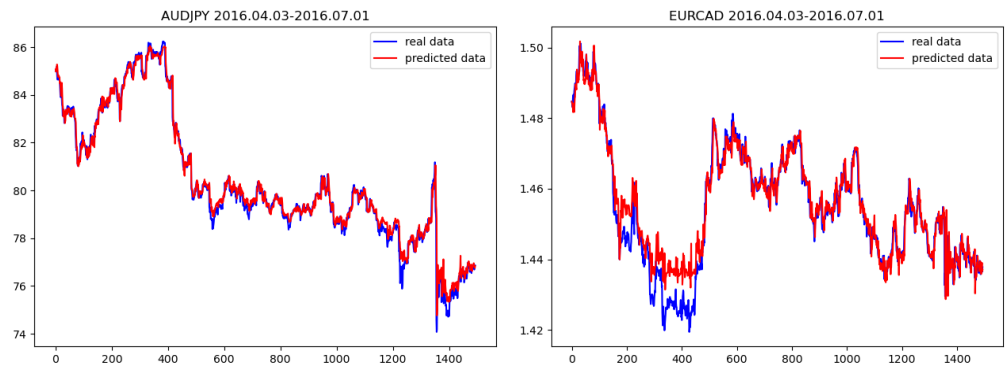

Fig 14 Total fitting curves of forex AUDJPY and forex EURCAD in 6 test periods

The fitting curves are shown in Fig.9. Tables 8 and 9 show the details of the trades where the six different trading models are evaluated by the 9-evaluation metrics shown in Section 3.2. As can be seen from tables, Model 6 has a more consistent performance in terms of returns than the other 5 models that have fluctuating returns. Compared to the first four models, Model 6 can make more profit with fewer trades. Comparing the P/L of strategy 5 and Model 6, the investment profitability of Model 6 is greater.

Table 8 nine performance metrics comparisons of forex AUDJPY with 6 trading models

| M | TP | PR | SUMPR | TN | TN+ | TN- | Avg_return | Avg_profit | Avg_loss | P/L | Accuracy |
|---|---|---|---|---|---|---|---|---|---|---|---|
| 1 | 1 | 1.62% | 1.84% | 229 | 121 | 108 | 0.007 | 0.153 | -0.156 | 0.979 | 52.84% |
|   | 2 | 2.77% |  | 252 | 117 | 135 | 0.011 | 0.155 | -0.114 | 1.362 | 46.43% |
|   | 3 | -3.18% |  | 252 | 120 | 132 | -0.013 | 0.119 | -0.132 | 0.900 | 47.62% |
|   | 4 | -2.38% |  | 252 | 121 | 131 | -0.009 | 0.105 | -0.115 | 0.911 | 48.02% |
|   | 5 | -1.86% |  | 252 | 127 | 125 | -0.007 | 0.128 | -0.145 | 0.883 | 50.40% |
|   | 6 | 4.87% |  | 251 | 132 | 119 | 0.019 | 0.237 | -0.222 | 1.068 | 52.59% |
| 2 | 1 | 2.72% | -0.71% | 229 | 117 | 112 | 0.012 | 0.163 | -0.146 | 1.117 | 51.09% |
|   | 2 | -0.12% |  | 252 | 120 | 132 | 0.000 | 0.139 | -0.127 | 1.092 | 47.62% |
|   | 3 | -1.08% |  | 252 | 128 | 124 | -0.004 | 0.120 | -0.132 | 0.905 | 50.79% |
|   | 4 | -4.06% |  | 252 | 113 | 139 | -0.016 | 0.105 | -0.114 | 0.916 | 44.84% |
|   | 5 | -1.59% |  | 252 | 126 | 126 | -0.006 | 0.130 | -0.143 | 0.912 | 50.00% |
|   | 6 | 3.42% |  | 251 | 123 | 128 | 0.014 | 0.248 | -0.212 | 1.172 | 49.00% |
| 3 | 1 | 1.59% | 8.41% | 229 | 120 | 109 | 0.007 | 0.154 | -0.155 | 0.994 | 52.40% |
|   | 2 | -0.29% |  | 252 | 121 | 131 | -0.001 | 0.137 | -0.129 | 1.064 | 48.02% |
|   | 3 | 4.13% |  | 252 | 135 | 117 | 0.016 | 0.133 | -0.118 | 1.126 | 53.57% |
|   | 4 | -0.58% |  | 252 | 123 | 129 | -0.002 | 0.110 | -0.110 | 1.006 | 48.81% |
|   | 5 | 1.31% |  | 252 | 119 | 133 | 0.005 | 0.150 | -0.125 | 1.206 | 47.22% |
|   | 6 | 2.26% |  | 251 | 121 | 130 | 0.009 | 0.248 | -0.213 | 1.162 | 48.21% |
| 4 | 1 | -2.45% | -8.08% | 229 | 102 | 127 | -0.011 | 0.161 | -0.149 | 1.083 | 44.54% |
|   | 2 | -2.35% |  | 252 | 133 | 119 | -0.009 | 0.117 | -0.151 | 0.777 | 52.78% |
|   | 3 | 2.19% |  | 252 | 130 | 122 | 0.009 | 0.130 | -0.121 | 1.077 | 51.59% |
|   | 4 | -1.72% |  | 252 | 119 | 133 | -0.007 | 0.109 | -0.111 | 0.987 | 47.22% |
|   | 5 | -0.86% |  | 252 | 118 | 134 | -0.003 | 0.142 | -0.132 | 1.080 | 46.83% |
|   | 6 | -2.88% |  | 251 | 135 | 116 | -0.011 | 0.202 | -0.260 | 0.777 | 53.78% |
| 5 | 1 | 0.53% | 7.29% | 127 | 67 | 60 | 0.004 | 0.162 | -0.172 | 0.941 | 52.76% |
|   | 2 | 2.39% |  | 108 | 54 | 54 | 0.022 | 0.164 | -0.120 | 1.369 | 50.00% |
|   | 3 | -0.44% |  | 96 | 48 | 48 | -0.005 | 0.117 | -0.126 | 0.927 | 50.00% |
|   | 4 | -0.12% |  | 86 | 43 | 43 | -0.001 | 0.135 | -0.138 | 0.979 | 50.00% |
|   | 5 | 0.46% |  | 55 | 28 | 27 | 0.008 | 0.130 | -0.118 | 1.104 | 50.91% |
|   | 6 | 4.47% |  | 85 | 46 | 39 | **0.053** | **0.353** | **-0.302** | 1.170 | 54.12% |
| 6 | 1 | 0.1% | **8.81%** | 125 | 59 | 66 | 0.001 | 0.175 | -0.155 | 1.130 | 47.20% |
|   | 2 | 0.26% |  | 104 | 47 | 57 | 0.003 | 0.139 | -0.110 | 1.262 | 45.19% |
|   | 3 | 2.50% |  | 113 | 63 | 50 | 0.022 | 0.097 | -0.072 | 1.348 | 55.75% |
|   | 4 | 1.51% |  | 93 | 54 | 39 | 0.016 | 0.130 | -0.142 | 0.920 | **58.06%** |
|   | 5 | 0.39% |  | 92 | 39 | 53 | 0.004 | 0.177 | -0.123 | **1.441** | 42.39% |
|   | 6 | 4.05% |  | 148 | 72 | 76 | 0.027 | 0.294 | -0.225 | 1.305 | 48.65% |

Table 9 nine performance metrics comparisons of forex EURCAD with 6 trading models

| M | TP | PR | SUMPR | TN | TN+ | TN- | Avg_return | Avg_profit | Avg_loss | P/L | Accuracy |
|---|---|---|---|---|---|---|---|---|---|---|---|
| 1 | 1 | 3.59% | 3.44% | 229 | 119 | 110 | 0.016 | 0.110 | -0.087 | 1.273 | 51.97% |
|   | 2 | -3.14% |  | 252 | 111 | 141 | -0.012 | 0.087 | -0.091 | 0.958 | 44.05% |
|   | 3 | 0.05% |  | 252 | 118 | 134 | 0.000 | 0.088 | -0.077 | 1.141 | 46.83% |
|   | 4 | -1.81% |  | 252 | 115 | 137 | -0.007 | 0.078 | -0.079 | 0.991 | 45.63% |
|   | 5 | 3.81% |  | 252 | 136 | 116 | 0.015 | 0.089 | -0.071 | 1.247 | 53.97% |
|   | 6 | 0.94% |  | 251 | 126 | 125 | 0.004 | 0.112 | -0.106 | 1.062 | 50.20% |
| 2 | 1 | 1.38% | 4.5% | 229 | 120 | 109 | 0.006 | 0.100 | -0.098 | 1.027 | 52.40% |
|   | 2 | -1.28% |  | 252 | 125 | 127 | -0.005 | 0.085 | -0.093 | 0.906 | 49.60% |
|   | 3 | -0.4% |  | 252 | 124 | 128 | -0.002 | 0.082 | -0.082 | 0.993 | 49.21% |
|   | 4 | -0.3% |  | 252 | 119 | 133 | -0.119 | 0.082 | -0.076 | 1.085 | 47.22% |
|   | 5 | 4.14% |        | 252 | 136 | 116 | 0.016  | 0.090 | **-0.070** | 1.290 | 53.97% |
|   | 6 | 0.96% |        | 251 | 121 | 130 | 0.004  | 0.117 | -0.102 | 1.152 | 48.21% |
| 3 | 1 | 1.24% | 3.74%  | 103 | 54  | 49  | 0.012  | 0.121 | -0.108 | 1.120 | 52.43% |
|   | 2 | -0.19%|        | 56  | 26  | 30  | -0.003 | 0.117 | -0.108 | 1.085 | 46.43% |
|   | 3 | 0.37% |        | 128 | 60  | 68  | 0.003  | 0.099 | -0.082 | 1.209 | 46.88% |
|   | 4 | -0.78%|        | 95  | 43  | 52  | -0.008 | 0.082 | -0.083 | 0.991 | 45.26% |
|   | 5 | 1.28% |        | 159 | 84  | 75  | 0.008  | 0.094 | -0.088 | 1.065 | 52.83% |
|   | 6 | 1.82% |        | 159 | 76  | 83  | 0.011  | **0.143** | -0.109 | 1.311 | 47.80% |
| 4 | 1 | 3.50% | -0.25% | 229 | 126 | 103 | 0.015  | 0.104 | -0.093 | 1.116 | 55.02% |
|   | 2 | -2.62%|        | 252 | 121 | 131 | -0.010 | 0.082 | -0.096 | 0.856 | 48.02% |
|   | 3 | -0.47%|        | 252 | 124 | 128 | -0.002 | 0.081 | -0.083 | 0.986 | 49.21% |
|   | 4 | -2.17%|        | 252 | 116 | 136 | -0.009 | 0.076 | -0.081 | 0.941 | 46.03% |
|   | 5 | 1.67% |        | 252 | 134 | 118 | 0.007  | 0.082 | -0.079 | 1.038 | 53.17% |
|   | 6 | -0.15%|        | 251 | 120 | 131 | -0.001 | 0.113 | -0.105 | 1.080 | 47.81% |
| 5 | 1 | 4.27% | 5.06%  | 133 | 74  | 59  | -0.042 | 0.133 | -0.095 | **1.406** | 55.64% |
|   | 2 | -0.40%|        | 54  | 26  | 28  | -0.039 | 0.111 | -0.117 | 0.947 | 48.15% |
|   | 3 | -1.75%|        | 112 | 48  | 64  | **0.066** | 0.085 | -0.091 | 0.934 | 42.86% |
|   | 4 | 1.13% |        | 124 | 62  | 62  | -0.031 | 0.097 | -0.079 | 1.231 | 50.00% |
|   | 5 | 1.71% |        | 151 | 84  | 67  | -0.161 | 0.082 | -0.077 | 1.060 | 55.63% |
|   | 6 | 0.22% |        | 147 | 69  | 78  | 0.018  | 0.135 | -0.117 | 1.158 | 46.94% |
| 6 | 1 | 4.31% | **5.65%** | 144 | 86 | 58 | 0.030 | 0.119 | -0.102 | 1.167 | **59.72%** |
|   | 2 | 2.11% |        | 150 | 74  | 76  | 0.014  | 0.074 | -0.100 | 0.742 | 49.33% |
|   | 3 | 0.48% |        | 152 | 77  | 75  | 0.003  | 0.086 | -0.082 | 1.050 | 50.66% |
|   | 4 | 0.1%  |        | 131 | 60  | 71  | 0.001  | 0.091 | -0.078 | 1.162 | 45.80% |
|   | 5 | 2.14% |        | 161 | 84  | 77  | 0.013  | 0.099 | -0.080 | 1.235 | 52.17% |
|   | 6 | 0.94% |        | 166 | 77  | 89  | 0.006  | 0.141 | -0.111 | 1.266 | 46.39% |

## 4. Conclusions

In this article, a novel learning model using deep fuzzy SOM companied with GRU networks is proposed, based on which an intelligent trading system is constructed. In order to recognize fluctuation patterns of financial data, the images of extended candlestick charts with both prices and volumes are constructed. Subsequently, to capture the features of fluctuation patterns in different time scales, the two-layer deep fuzzy self-organizing map is implemented for clustering of the candlestick patterns in an unsupervised learning way. Finally, for different patterns obtained by clustering, GRU networks are trained to implement the prediction tasks and financial trading decisions can be made. By using real financial dataset including commodity futures and forex, various experiments are implemented to verify the effectiveness of the proposed models. We hope that the results of this work can enrich the forecasting methods of financial time series and provide some new ideas for the quantitative trading.

# Appendix

Table 10 Nine performance metrics for DFSOM in the whole China's commodity futures dataset

| Code | PR | TN | TN+ | TN- | Avg_return | Avg_profit | Avg_loss | P/L | Accuracy |
|---|---|---|---|---|---|---|---|---|---|
| AG | 55.77% | 1976 | 1020 | 956 | 0.028 | 2.822 | 0.318 | 1.11 | 51.62% |
| B | 47.85% | 1480 | 768 | 712 | 0.032 | 3.233 | 0.288 | 1.14 | 51.89% |
| BU | 16.52% | 1420 | 716 | 704 | 0.012 | 1.163 | 0.428 | 1.04 | 50.42% |
| C | 19.16% | 1294 | 657 | 637 | 0.015 | 1.481 | 0.180 | 1.13 | 50.77% |
| EB | 62.00% | 1139 | 607 | 532 | 0.054 | 5.443 | 0.337 | 1.18 | 53.29% |
| FG | 23.96% | 1177 | 603 | 574 | 0.020 | 2.036 | 0.326 | 1.07 | 51.23% |
| HC | 17.97% | 1037 | 564 | 473 | 0.017 | 1.733 | 0.237 | 0.97 | 54.39% |
| JM | 28.16% | 1446 | 742 | 704 | 0.019 | 1.947 | 0.260 | 1.25 | 51.31% |
| MA | 25.93% | 1408 | 729 | 679 | 0.018 | 1.842 | 0.336 | 1.04 | 51.78% |
| PB | 21.77% | 1964 | 1039 | 925 | 0.011 | 1.108 | 0.123 | 1.83 | 52.90% |
| PP | 20.25% | 1484 | 769 | 715 | 0.014 | 1.365 | 0.268 | 1.03 | 51.82% |
| RB | 17.91% | 1428 | 759 | 669 | 0.013 | 1.254 | 0.233 | 0.98 | 53.15% |
| RM | 39.32% | 1438 | 770 | 668 | 0.027 | 2.734 | 0.212 | 1.11 | 53.55% |
| RR | 15.03% | 1226 | 660 | 566 | 0.012 | 1.226 | 0.108 | 1.22 | 53.83% |
| SA | 18.88% | 1326 | 683 | 643 | 0.014 | 1.424 | 0.339 | 1.02 | 51.51% |
| SC | 60.46% | 2468 | 1267 | 1201 | 0.024 | 2.450 | 0.346 | 1.09 | 51.34% |
| SN | 21.76% | 2071 | 1081 | 990 | 0.011 | 1.051 | 0.197 | 1.02 | 52.20% |
| SP | 14.03% | 1330 | 682 | 648 | 0.011 | 1.055 | 1.743 | 0.12 | 51.28% |
| SR | 19.61% | 1465 | 754 | 711 | 0.013 | 1.339 | 0.194 | 1.08 | 51.47% |
| SS | 32.60% | 1914 | 1020 | 894 | 0.017 | 1.703 | 0.207 | 0.98 | 53.29% |
| TA | 27.50% | 1409 | 723 | 686 | 0.020 | 1.952 | 0.271 | 1.09 | 51.31% |
| Y | 20.88% | 1454 | 747 | 707 | 0.014 | 1.436 | 0.302 | 1.04 | 51.38% |

Table 11 Nine performance metrics for DFSOM in the whole foreign exchange dataset

| Code | PR | TN | TN+ | TN- | Avg_return | Avg_profit | Avg_loss | P/L | Accuracy |
|---|---|---|---|---|---|---|---|---|---|
| AUDJPY | 23.39% | 3790 | 2023 | 1767 | 0.006 | 0.122 | 0.093 | 1.15 | 53.38% |
| AUDUSD | 17.84% | 2797 | 1430 | 1367 | 0.006 | 0.102 | 0.089 | 1.09 | 51.13% |
| EURCAD | 10.94% | 2301 | 1166 | 1135 | 0.005 | 0.094 | 0.084 | 1.08 | 50.67% |
| EURCHF | 7.78% | 4401 | 2286 | 2115 | 0.002 | 0.043 | 0.040 | 1.00 | 51.94% |
| EURGBP | 18.02% | 3660 | 1881 | 1779 | 0.005 | 0.093 | 0.084 | 1.05 | 51.39% |
| EURJPY | 13.48% | 4415 | 2209 | 2206 | 0.003 | 0.102 | 0.096 | 1.06 | 50.03% |
| EURUSD | 7.90% | 881 | 453 | 428 | 0.009 | 0.101 | 0.084 | 1.14 | 51.42% |
| GBPCHF | 21.46% | 4415 | 2293 | 2122 | 0.005 | 0.100 | 0.090 | 1.02 | 51.94% |
| GBPJPY | 21.29% | 4415 | 2214 | 2201 | 0.005 | 0.144 | 0.124 | 1.16 | 50.15% |
| GBPUSD | 20.06% | 4415 | 2234 | 2181 | 0.005 | 0.104 | 0.095 | 1.07 | 50.60% |
| NZDJPY | 11.38% | 4414 | 2205 | 2209 | 0.003 | 0.124 | 0.118 | 1.05 | 49.95% |
| NZDUSD | 15.16% | 4160 | 2112 | 2048 | 0.004 | 0.108 | 0.100 | 1.04 | 50.77% |
| USDCAD | 12.25% | 4163 | 2102 | 2061 | 0.003 | 0.081 | 0.075 | 1.06 | 50.49% |
| USDCHF | 7.30% | 4176 | 2074 | 2102 | 0.002 | 0.068 | 0.065 | 1.07 | 49.66% |
| USDJPY | 14.75% | 4413 | 2177 | 2236 | 0.003 | 0.103 | 0.097 | 1.10 | 49.33% |
| XAGUSD | 32.52% | 3653 | 1903 | 1750 | 0.009 | 0.226 | 0.209 | 0.99 | 52.09% |
| XAUUSD | 15.59% | 4184 | 2089 | 2095 | 0.004 | 0.130 | 0.123 | 1.06 | 49.93% |

**Acknowledgements** — This research is supported by the National Natural Science Foundation of China (Nos: 61402267); Shandong Provincial Natural Science Foundation (ZR2019MF020).